\documentclass{article} 
\usepackage{iclr2027_conference,times}
\usepackage{stmaryrd}
\usepackage{times}
\usepackage{latexsym}
\usepackage{amsmath}
\usepackage{lineno}
\usepackage{amssymb}
\usepackage{mathtools}
\usepackage{amsthm}
\usepackage{microtype}
\usepackage{inconsolata}
\usepackage{setspace}
\usepackage{graphicx} 
\usepackage{enumitem}
\usepackage{subcaption}
\usepackage{booktabs}
\usepackage{multirow, makecell, threeparttable, tabularx}
\usepackage{array}
\usepackage[T1]{fontenc}
\usepackage[dvipsnames, table]{xcolor}
\usepackage{listings}
\usepackage{xspace}
\usepackage{etoc}
\usepackage[most]{tcolorbox}
\usetikzlibrary{patterns,patterns.meta}
\usepackage{caption}
\usepackage{mfirstuc}

\usepackage[normalem]{ulem}

\usepackage[hidelinks]{hyperref}
\newcommand{\Tool}{Doc2Spec\xspace}
\newcommand{\beforecaption}{\vspace{-.05cm}\begin{spacing}{0.85}}
\newcommand{\aftercaption}{\vspace{-.15cm}\end{spacing}}
\newcommand{\mycaption}[3]{\beforecaption\caption{\label{#1}{#2} #3}\aftercaption}

\newcommand{\eg}{\textit{e.g.}}

\newcommand{\sxia}[1]{{\color{orange} \sf (SH: #1)}}

\newcommand{\boldparagraph}[1]{\noindent{\textbf{#1}}}

\newcommand{\stepname}[1]{%
  \textsf{\textit{\capitalisewords{#1}}}%
}

\newcommand{\NA}{\cellcolor{gray!20}}
\definecolor{CodeBg}{RGB}{248,248,252}
\definecolor{CodeFrame}{RGB}{220,220,230}

\lstdefinestyle{CodeBase}{
  basicstyle=\ttfamily\scriptsize,
  backgroundcolor=\color{CodeBg},
  frame=single,
  rulecolor=\color{CodeFrame},
  numbers=left,
  numberstyle=\tiny\color{gray},
  numbersep=8pt,
  showstringspaces=false,
  tabsize=2,
  breaklines=true,
  breakatwhitespace=true,
  breakautoindent=true
  keepspaces=true,
  xleftmargin=.15in,
  framexleftmargin=.15in
}

\lstdefinestyle{CodePrompt}{
  style=CodeBase,
  stringstyle=\color{RoyalBlue},   
  commentstyle=\color{Gray},
  numbers=none
}

\lstdefinelanguage{appendixjson}{
  basicstyle=\small,
  backgroundcolor=\color{CodeBg},
  showstringspaces=false,
  breaklines=true,
  columns=fullflexible,
  morestring=[b]",
  stringstyle=,
  comment=[l]{//},
  morecomment=[s]{/*}{*/},
  literate=
   *{0}{{0}}1 {1}{{1}}1 {2}{{2}}1 {3}{{3}}1 {4}{{4}}1
    {5}{{5}}1 {6}{{6}}1 {7}{{7}}1 {8}{{8}}1 {9}{{9}}1
    {:}{{:}}1 {,}{{,}}1 {\{}{{\{}}1 {\}}{{\}}}1 {[}{{[}}1 {]}{{]}}1
}

\lstdefinestyle{llmtext}{
  basicstyle=\ttfamily\footnotesize,
  columns=fullflexible,
  breaklines=true,
  breakatwhitespace=true,
  keepspaces=true,
  showstringspaces=false,
  tabsize=2,
  escapeinside={(*@}{@*)},
  breakindent=0pt,
  breakautoindent=false,
  columns=fullflexible,
  keepspaces=true
}

\newcommand{\hlline}[1]{%
  \rlap{%
    \color{yellow!50}%
    \rule[-0.3\baselineskip]{#1\linewidth}{\baselineskip}%
  }%
}

\lstdefinelanguage{Solidity}{
    sensitive=true,
    morecomment=[l]{//},
    morecomment=[s]{/*}{*/},
    morestring=[b]"
}

\lstdefinestyle{SolidityStyle}{
    language=Solidity,
    basicstyle=\ttfamily\scriptsize,
    frame=single,
    numbers=left,
    numberstyle=\tiny,
    breaklines=true,
    xleftmargin=2pt,
    xrightmargin=2pt,
    captionpos=b,
    keywordstyle=\normalfont,
    commentstyle=\normalfont,
    stringstyle=\normalfont,
    morekeywords=[1]{public},
    keywordstyle=[1]{\color{blue}\bfseries},
    morekeywords=[2]{returns},
    keywordstyle=[2]{\color{RedViolet}},
    morekeywords=[3]{uint256, bool},
    keywordstyle=[3]{\color{teal}},
    rulecolor=\color{black},   
    moredelim=[is][\color{orange}\bfseries]{|}{|}, 
    moredelim=[is][\color{blue}\bfseries]{~}{~}, 
    moredelim=[is][\color{teal}]{!}{!}, 
}

\lstdefinelanguage{Rust}{
    keywords={
        as, break, const, continue, crate, else, enum, extern, false, fn, for,
        if, impl, in, let, loop, match, mod, move, mut, pub, ref, return, self,
        Self, static, struct, super, trait, true, type, unsafe, use, where, while
    },
    morecomment=[l]{//},
    morecomment=[s]{/*}{*/},
    morestring=[b]",
    sensitive=true
}

\lstdefinestyle{RustStyle}{
    language=Rust,
    basicstyle=\ttfamily\scriptsize,
    keywordstyle=\color{blue}\bfseries,
    commentstyle=\color{gray}\itshape,
    stringstyle=\color{red},
    numbers=left,
    numberstyle=\tiny,
    breaklines=true,
    frame=single,
    xleftmargin=2pt,
    xrightmargin=2pt,
    captionpos=b,
    rulecolor=\color{black},   
    moredelim=[is][\color{orange}\bfseries]{|}{|}, 
}

\newtcolorbox{llmex}[2][]{%
  enhanced,
  breakable,
  colback=black!1,
  colframe=black!28,
  colbacktitle=black!5,
  coltitle=black!82,
  title={#2},
  fonttitle=\sffamily\bfseries,
  boxrule=0.55pt,
  arc=3pt,
  left=7pt,right=7pt,top=5pt,bottom=5pt,
  before skip=7pt,after skip=8pt,
  #1
}

\DeclareUnicodeCharacter{2460}{\ding{172}}
\DeclareUnicodeCharacter{2461}{\ding{173}}
\DeclareUnicodeCharacter{2462}{\ding{174}}
\DeclareUnicodeCharacter{2463}{\ding{175}}
\usepackage{promptpanels}
\title{Doc2Spec: Synthesizing Formal Programming Specifications from Natural Language via Grammar Induction}

\iclrfinalcopy

\author{
Shihao Xia \& Mengting He \\
The Pennsylvania State University
\And
Haomin Jia \& Xinyan Zhao \\
University of the Chinese Academy of Sciences
\AND
Linhai Song\thanks{Corresponding author: \texttt{songlinhai@ict.ac.cn}} \\
Institute of Computing Technology, Chinese Academy of Sciences
}

\begin{document}

\maketitle
\fancyhead{}

\begin{abstract}

Ensuring that API implementations and usage comply with natural language programming rules is critical for software correctness, security, and reliability. 
Formal verification can provide strong guarantees but requires precise 
specifications, which are difficult and costly to write manually. 
To address this challenge, we present \Tool{}, a multi-agent framework that automatically induces a domain-specific grammar from natural-language API rules and uses it to guide specification generation. 
\Tool{} fixes a domain-agnostic logical skeleton as a grammar template, prompts LLMs to infer domain-specific predicates and sorts, 
and formalizes each rule within the resulting grammar, turning an unreliable one-shot translation into a sequence of constrained, checkable steps. 
Across six benchmarks spanning Solidity and Rust, 
\Tool{} improves precision by $0.28$ and recall by $0.37$ over baselines that lack grammar induction or 
perform it in one unstaged step, 
demonstrating the benefits of grammar-guided formalization and the staged pipeline.
Moreover, its formalized rules enable symbolic-execution tools to uncover 142 previously unknown rule violations, confirming the rules' correctness and practical usefulness.

\end{abstract}

\section{Introduction}
\label{sec:intro}

Modern software systems are increasingly constructed by composing and reusing modules and libraries~\citep{mikkonen2019software, saied2018improving}.
These modules are often developed independently and later integrated into larger systems, making well-defined application programming interfaces (APIs) essential for correctness, security, and maintainability~\citep{lamothe2021systematic, rauf2019systematic}.
Beyond function signatures, APIs also specify semantic requirements for
correct implementation and proper use~\citep{de2017api, cummaudo2019should}. 
These requirements are usually written in natural language in 
API documentation or code comments and include constraints on inputs 
and outputs, program state changes, and error handling~\citep{pandita2012inferring, zhang2021discovering}.

Violating API requirements can cause serious security and reliability issues. 
For example, the ERC20 standard in Figure~\ref{fig:overview} specifies rules for implementing public smart contract APIs~\citep{erc20}. 
A violation, such as failing to check caller privileges in \texttt{transferFrom()} 
as required by the highlighted sentence, can let attackers steal digital assets. 
Similar problems occur 
in other systems, such as Rust’s \texttt{unsafe} APIs, where 
incorrect use can lead to safety or correctness violations (\eg, 
out-of-bounds, invalid free)~\citep{qin2020understanding}. 
Ensuring that API implementations and their usage follow the documented natural-language rules is therefore critical.

Formal verification is a principled way to ensure rule compliance, 
using techniques such as symbolic execution~\citep{asadian2022applying, asadian2024monitor}, 
model checking~\citep{nehai2018model, martinelli2019model}, and deductive verification~\citep{summers2020automating, anureev2019two}. 
These methods provide mathematically grounded guarantees that an implementation satisfies a specification. However, they require precise formal specifications from developers to describe the intended code behaviors. 
Writing such specifications is difficult, as it demands expertise in formal methods, 
deep knowledge of the code, and careful treatment of corner cases. 
As a result, despite their strength, formal verification techniques are still hard to adopt in practice~\citep{mazzara2020towards, reid2020towards}.

\begin{figure*}[t]
\centering
    \includegraphics[width=\textwidth]{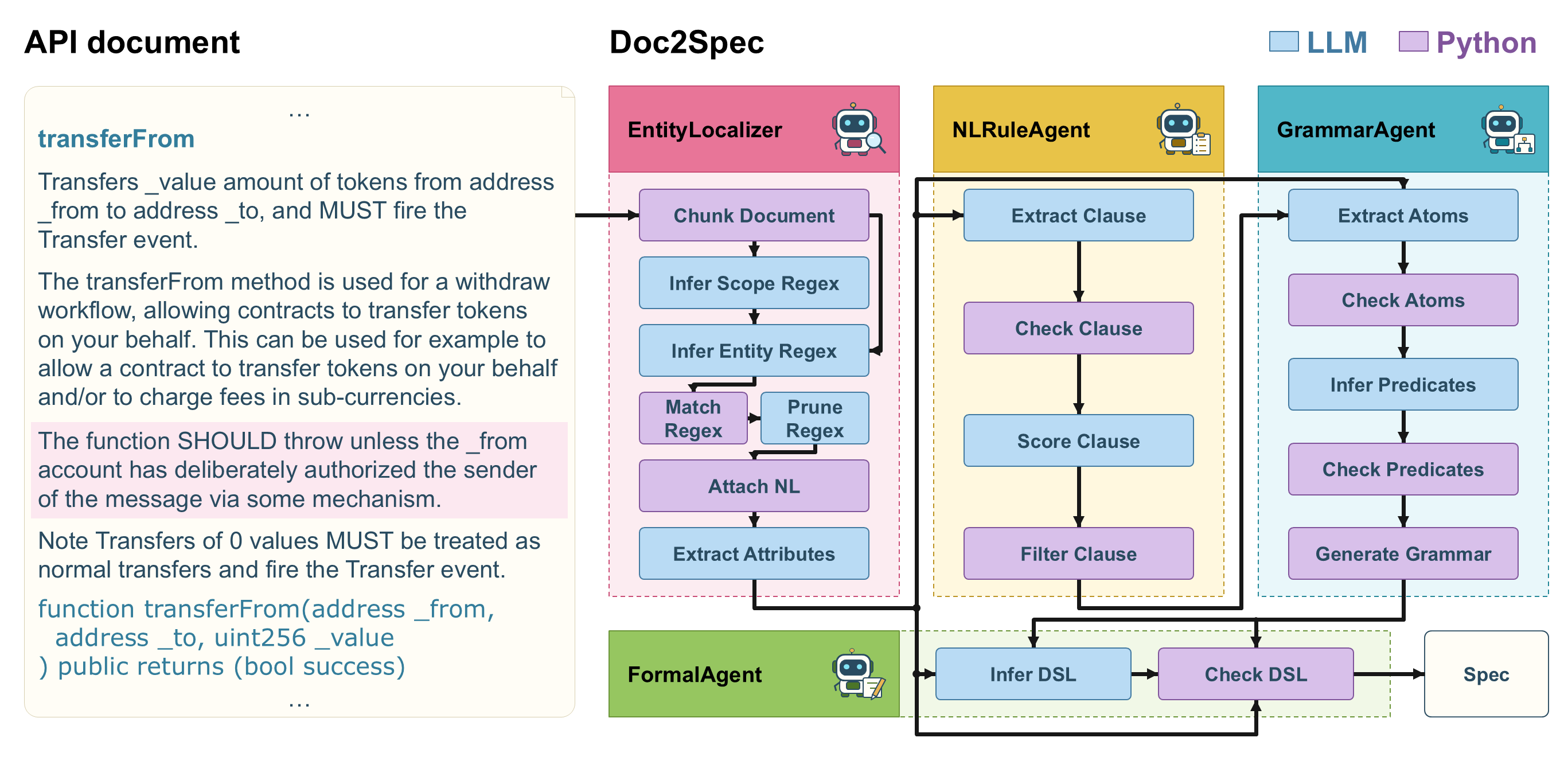}
    \vspace{-0.1in}
    \mycaption{fig:overview}{Workflow of \Tool{}.}
    {\textit{(LLM-based components are in blue, 
    and Python tools are in purple. Example outputs of the four agents are shown in Figures~\ref{fig:agent-output-entity}--\ref{fig:agent-output-formal}
    in the appendix.)}}
\vspace{-0.1in}
\end{figure*}

To address this challenge, recent work has explored using large 
language models (LLMs) to automatically synthesize formal specifications 
from natural language. At a high level, this task can be viewed as a translation 
problem, and LLMs have shown strong ability to map between different semantic representations~\citep{zhu2024multilingual, schneider2024evaluating}. Existing approaches generally fall into two categories. 
The first uses in-context learning, where carefully designed prompts ~\citep{shi2023large,wei2022chain}
or multi-agent pipelines~\citep{zhang2024offline, zhang2025which} guide LLMs to generate specifications~\citep{neurosymbol-1, wu2022autoformalization,misu2024towards, Logic-LM, Logic-LM++,yao2023leveraging, kamath2023finding, wu2024lemur, sun2025classinvgen, mugnier2025laurel, loughridge2025dafnybench, LINC, wen2024enchanting, SpecGen, ryu_divide_2025, AutoVerus, VeriStruct, sun2024clover}. 
Although effective in some settings, these methods often require substantial domain-specific customization and do not reliably generalize to 
new or generic scenarios. 
The second category relies on post-training 
or fine-tuning LLMs~\citep{hahn2023formal, chen2023nl2tl, SAFE, azerbayev2023llemma,yang2023leandojo, zrelli2024advancing, polu2023formal,first2023baldur, chakraborty2025towards}. However, in many cases, the required labeled data is scarce, costly to obtain, or unavailable.

This raises a fundamental question: how can we build a general technique to translate natural language programming 
rules into formal specifications across diverse coding scenarios, 
without relying on post-training or domain-specific customization?

As shown in Figure~\ref{fig:overview}, 
we address this problem by inducing a grammar of formal specifications from the natural-language rules 
and using it to guide specification generation. 
Our key insight is that a formal specification rests on three parts: a general logical skeleton shared across domains (conditions, obligations, and how they combine), a domain-specific vocabulary of predicates and sorts, 
and the content of each individual specification. 
Rather than producing all three at once, our approach builds them in order: each part constrains the next, so every LLM step works within a fixed space rather than an unconstrained one. 
Because the intermediate results are explicit and structured, each step can also be checked deterministically, turning an unreliable one-shot translation into a sequence of verifiable steps.

Building on this insight, we propose \Tool{}, 
a multi-agent framework that takes natural-language documents of programming 
rules as input and generates specifications suitable for formal verification 
(\eg, symbolic execution). \Tool{} operates in four stages: identifying constrained entities, extracting rule-related sentences, inducing a grammar from them, and using the grammar to formalize the rules. 
The key step is grammar induction: we design a domain-agnostic grammar template and prompt an LLM to summarize the domain-specific predicates and sorts from the natural-language rules, 
which together with the template form a domain-specific grammar. 
Unlike CLOVER~\citep{ryu_divide_2025}, which extracts predicates but relies on a fixed set of sorts, \Tool{} infers sorts dynamically for each domain, 
making it more flexible across domains.

We evaluate \Tool{} on six documents of natural-language rules, covering both API implementation and API usage and spanning two programming languages (Solidity and Rust).
We compare \Tool{} with three baselines: 
one that formalizes rules in a single prompt, 
one that induces the grammar in a single prompt and then uses the grammar to formalize rules, 
and one that relies on a manually crafted grammar for rule formalization~\citep{SymGPT}. 
\Tool{} improves precision by 0.28 and recall by 0.37 over the first two baselines, highlighting the benefits of grammar-guided formalization and its staged agent pipeline. 
Compared with the third baseline, \Tool{} achieves higher recall on two benchmarks despite requiring no manual grammar engineering. 
Moreover, we integrate the formalized rules with downstream symbolic-execution tools: they uncover 142 previously unknown rule violations in real-world Solidity smart contracts and open-source Rust crates, with only 21 false alarms, 
demonstrating both the correctness and the practical utility of the generated specifications.

In sum, we make the following contributions:

 \begin{itemize}
 [leftmargin=*, itemsep=2pt, topsep=2pt]
   \item We propose a novel approach for automatically inducing grammars from natural language rules.
   
   \item We design \Tool{}, a multi-agent system that induces grammars from API rules and uses them to translate the rules into formal specifications for verification.
   
   \item We evaluate \Tool{} on six benchmarks, demonstrating its effectiveness and the practical utility of its formalized rules.
 \end{itemize}

\section{Background and Related Work}
\label{sec:related}

This section provides background on this project, including natural-language API usage and implementation rules, 
as well as prior work on using LLMs to automate formal verification.

\subsection{Natural Language API Rules}

Modules and libraries expose APIs that allow external code to 
access their functionality. These APIs are typically accompanied 
by documentation or comments that describe how to use 
them correctly, including valid inputs, call order, return-value checks, and calling context.
For example, comments before the declaration of Rust’s standard library function \texttt{dealloc(ptr: *mut u8, layout: Layout)} specify 
that \texttt{ptr} cannot be \texttt{NULL}. Otherwise, 
using the function with a \texttt{NULL} pointer introduces 
a serious security vulnerability.

Some APIs also serve as public standards and are implemented 
by multiple groups in different versions. In such cases, the API 
documentation specifies mandatory requirements that 
implementations must follow to ensure consistent and expected behavior across versions.
Figure~\ref{fig:overview} illustrates this scenario using ERC20, 
a public standard for Solidity smart contracts that manage digital tokens 
on Ethereum. Once a contract is deployed, anyone can 
invoke its public function \texttt{transferFrom()}. Therefore, the 
function must verify the caller’s privileges as required by the highlighted sentence. Otherwise, any user could steal tokens by using this function to transfer tokens to his own address.

\subsection{Automated Formal Verification via LLMs}

Existing techniques that use LLMs to automate or assist with formal software verification mainly focus on generating verification specifications and proofs.
Traditionally, these artifacts were written manually, which is tedious and requires substantial expertise in both formal 
verification and the software’s domain. Broadly, existing techniques can be divided into two groups.

First, some researchers treat LLMs as black boxes and apply in-context learning to generate specifications and proofs. 
Various strategies have been proposed to improve performance, 
including few-shot prompting~\citep{neurosymbol-1,wu2022autoformalization}, chain-of-thought prompting~\citep{misu2024towards},
feedback from prover error messages~\citep{Logic-LM,Logic-LM++,yao2023leveraging, kamath2023finding, wu2024lemur, sun2025classinvgen, mugnier2025laurel, loughridge2025dafnybench}, majority voting over multiple specification generations 
for the same input~\citep{LINC}, 
iterative generation until prover acceptance~\citep{wen2024enchanting}, 
structure-first specification synthesis~\citep{ryu_divide_2025},
mutation and ranking of erroneous LLM-generated specifications~\citep{SpecGen}, 
summarizing common mistakes and refining generation errors via a multi-agent system~\citep{AutoVerus, VeriStruct}, and joint generation of code and specifications with cross-checking~\citep{sun2024clover}.
While effective, these methods are often domain-specific and rely heavily on carefully engineered prompts and examples, limiting their generalization to new domains.

Second, other researchers post-train or fine-tune open-source LLMs for formal
verification tasks, achieving success on Verus~\citep{SAFE}, Lean~\citep{azerbayev2023llemma,yang2023leandojo,polu2023formal}, regular expressions~\citep{hahn2023formal}, 
temporal logic~\citep{chen2023nl2tl}, computation tree logic~\citep{zrelli2024advancing}, Isabelle/HOL~\citep{first2023baldur},
and F*~\citep{chakraborty2025towards}.
However, these techniques require substantial 
amounts of paired natural-language descriptions and corresponding specifications or proofs. Such labeled data are 
scarce in many domains~\citep{SAFE,chen2023nl2tl}, limiting the feasibility of fine-tuning open-source LLMs in practice.

\section{Problem Formulation}
\label{sec:problem}

We study the problem of extracting code entities, their usage or implementation rules, and a formal rule grammar 
from unstructured natural language documents.

Let $D = \{l_1, l_2, \ldots, l_n\}$ denote a natural language document consisting 
of text lines.
We assume no prior knowledge of the document’s domain, structure, or programming content. Given $D$, \Tool{} identifies a set of \emph{entities}
$\mathcal{T} = \{t_1, t_2, \ldots, t_m\}$,
where each entity corresponds to an API,
struct, or similar code element, and may have multiple attributes. 
\Tool{} also induces a rule grammar $\mathcal{G}$ 
in Extended Backus--Naur form (EBNF) and extracts a set of rules
$\mathcal{R} = \{r_1, r_2, \ldots, r_k\}$, 
where each rule applies to a specific entity and is valid under $\mathcal{G}$.
We formalize the problem as learning a mapping 
\[
f : D \longrightarrow (\mathcal{T}, \mathcal{G}, \mathcal{R}).
\]

For instance, when processing the document in Figure~\ref{fig:overview}, 
\Tool{} induces a grammar for it. 
It identifies the function \texttt{transferFrom()} 
in the document as a program entity, 
extracts the natural-language rules constraining it (including the highlighted sentence), 
and formalizes these rules according to the induced grammar.

\section{Design of \Tool{}}
\label{sec:meth}

This section presents the detailed design of \Tool{}, which takes an API document as input and operates in four stages: 
(1) locating code entities, 
(2) identifying natural language statements that specify usage or implementation rules,
(3) inducing a grammar in EBNF from the rules, 
and (4) formalizing the rules using the grammar.

Figure~\ref{fig:overview} shows the overall architecture of \Tool{}. 
\Tool{} is domain-agnostic: neither its pipeline nor its prompt design assumes any domain knowledge. 
Even the grammar begins as a basic template that captures only general logical structure, 
which is then filled with domain-specific predicates and sorts inferred automatically from the natural-language rules in the input document. 
Moreover, \Tool{} decomposes the task top-down, from document to rules, 
and then from each rule to atoms and predicates. 
Because each step performs a single, simple operation, 
a deterministic Python tool can verify its output before the next step begins (\eg, checking that the extracted text is faithful to the document or that all inputs have been processed), 
thereby improving overall correctness.
The appendix contains all prompt templates for Figure~\ref{fig:overview}
and the example LLM responses.

\subsection{Stage 1: Localizing Programming Entities}
\label{sec:entity}


\emph{EntityLocalizer} identifies the programming entities in each input document, attaches its natural-language description to each entity, 
and extracts its declaration attributes.
As shown in Figure~\ref{fig:overview}, it combines four
LLM passes with three lightweight Python tools
for data preprocessing and output validation.
%
%
EntityLocalizer's design addresses three key challenges.

First, an input document may be too long for an LLM to process in one prompt, 
and even when it fits, directly enumerating entities can be inconsistent: the same LLM may identify an entity at one occurrence 
but miss it at another.
To address this, we split each document into 80-line chunks 
(\stepname{Chunk Document}). For each chunk, we ask the LLM to 
infer regular expressions that match the first line of 
an entity declaration (\stepname{Infer Entity Regex}). 
A Python tool then applies these regular expressions to the chunk, ensuring that matching entities are extracted consistently (\stepname{Match Regex}).

Second, programs are hierarchical, and functions 
in different scopes (\eg, interfaces, structs, and classes) may have the same name. Without identifying an entity's scope, downstream verification tools cannot determine 
which entity a generated formal rule belongs to.
To address this, we use a separate LLM pass to infer scope patterns (\stepname{Infer Scope Regex}):
a regular expression matching the first line of a scope and capturing its name, together with the delimiters that open and close its body.
We apply these patterns to the document and track the 
delimiters to determine the line range of each scope. 
This allows us to place each matched entity in its scope chain.
We infer scope patterns before entity patterns. We provide the scope patterns, along with examples, to the entity pass. This helps the LLM distinguish scope syntax 
from entity syntax.

Third, inferred entity patterns may be incorrect or may match lines 
that are not entities governed by natural-language rules (\eg, import statements, 
code quoted in comments). 
We therefore validate the patterns in two steps. (1) A Python tool (\stepname{Match Regex}) 
checks that each pattern compiles and matches its example, 
and then finds all matching lines. (2) The LLM reviews each pattern and all of its matches and decides whether to keep it (\stepname{Prune Regex}). 
We keep a pattern if at least one match is a genuine entity declaration and discard it only if all matches are irrelevant.

After validating the entity patterns, a Python tool (\stepname{Attach NL}) attaches the natural-language block immediately preceding each identified 
entity, following the common convention that an entity's description appears 
before its declaration. Finally, an LLM pass (\stepname{Extract Attributes}) extracts the entity's attributes 
from its complete declaration and description, 
including its kind, name, ordered parameters and their types, return values, modifiers, visibility, enclosing scopes, and attached description, for subsequent agents. 
Figure~\ref{fig:agent-output-entity} (appendix) shows the function \texttt{transferFrom()} in Figure~\ref{fig:overview} extracted as an entity, together with its description and attributes.

\subsection{Stage 2: Extracting Natural-Language Rules}
\label{sec:nlrule}

\emph{NLRuleAgent} takes each entity produced by EntityLocalizer,
together with its declaration and attached natural-language description, 
and extracts from the description the sentences that state implementation or usage rules. 
As shown in Figure~\ref{fig:overview}, \emph{NLRuleAgent} contains two LLM passes with two Python tools.

The main challenge is that extracting rules directly from free-form 
text is unreliable in two ways. (1) The LLM may paraphrase, merge, 
or invent text instead of reporting what the document says. (2) The LLM may be inconsistent about which clauses count as rules\footnote{We call a text span that potentially states a rule a \emph{clause}, 
and a clause confirmed to state a rule a \emph{rule}.}. 
We address this by separating \emph{finding} candidate clauses 
from \emph{deciding} whether each is a rule, with a deterministic check in between.

First, the LLM extracts every clause in the input description that may state an independent rule for the entity (\stepname{Extract Clause}). 
As shown in Listing~\ref{lst:n1:template} in the appendix, 
the prompt requires each clause to express exactly one independently testable requirement: a sentence with multiple requirements is split into separate clauses, while a condition is kept with its consequence. 
The prompt also clarifies what does not count as a clause, excluding topic summaries, definitions, rationale, examples, and parameter descriptions. 
To reduce missed rules, the LLM must inspect every sentence, list item, and structured field in the input. 
Finally, it must copy each clause verbatim, which keeps the extracted clauses faithful to the document.

Second, a Python tool validates the returned clauses against the input description (\stepname{Check Clause}). It first normalizes both sides by lower-casing text, collapsing whitespace, standardizing quotation marks and dashes, 
and removing Markdown backticks. It then checks that each clause appears as a substring of the description. Clauses that the LLM altered or fabricated are discarded. The tool also removes duplicates and clauses that are fully contained in another retained clause.

Third, the LLM scores each remaining clause on whether it states a rule for the target entity (\stepname{Score Clause}). 
Each clause is evaluated independently twice ($i \in \{1, 2\}$),
producing a label $r_i$ ($r_i=1$ for a rule, $r_i=-1$ otherwise)
and a confidence score $c_i \in [0, 1]$.
As shown in Listing~\ref{lst:n3:template} in the appendix, the prompt defines 
what counts as a rule and which cases to exclude, following the same criteria as \stepname{Extract Clause}.
The prompt also emphasizes that the score indicates 
the LLM's confidence that the clause is a rule, 
rather than the strength or importance of the rule itself.

Finally, a Python tool computes a score using the following formula and treats a clause
with a score greater than 
$0.5$
as a rule (\stepname{Filter Clause}).
For example, Figure~\ref{fig:agent-output-rule} (appendix) shows the 
two confidence scores reported by the LLM for the highlighted sentence in Figure~\ref{fig:overview}, 
together with the score computed by the tool. Since the computed score exceeds 
$0.5$, the sentence is identified as a rule.

\[
\text{score}(s) \;=\; \frac{1}{1+\exp\!\left(-\sum_{i=1}^{2} r_i c_i\right)}.
\]

\subsection{Stage 3: Inducing Rule Grammar}
\label{sec:grammar}

\begin{figure}[t]
\centering
\begin{lstlisting}[
      style=CodePrompt,
      numbers=left,          % line numbers on the left
  numberstyle=\tiny,     % style for line numbers
  stepnumber=1,          % number every line
  numbersep=8pt,
    ]
<spec>    ::= <rule>+
<rule>    ::= [<modality>] <formula> [if <formula>] ;
<modality>::= MODALITY 
<formula> ::= <predicate> | not <formula> | (<formula>) | <formula> <op> <formula>
<op>      ::= and | or
<predicate> ::= PREDICATE([<sorts>])
<sorts>   ::= <sort> | <sort>, <sorts>
<sort>    ::= SORT | string | number | boolean | TargetAttr(string) | TargetReturn([number])


\end{lstlisting}
    \mycaption{fig:template}
{Simplified DSL Template.}
{\textit{(Nonterminals are enclosed in angle brackets, 
all-uppercase tokens are placeholders instantiated during grammar induction,
and all other tokens are terminals written literally. 
TargetAttr(string) selects an attribute of the target, and TargetReturn([number]) selects a return value of the target.)}}
\vspace{-0.1in}
\end{figure}

\emph{GrammarAgent} induces an EBNF grammar by filling 
in a fixed DSL template rather than generating one from scratch. 
We observe that most API rules state that a property holds under a condition, 
where both the property and the condition are expressed with domain-specific predicates. The condition is optional, since some rules apply unconditionally, 
and rules carry different levels of obligation (\eg, ``must'', ``may''), 
so we include a modality indicating whether a violation is an error or a warning. Figure~\ref{fig:template} shows the resulting template. 
It fixes this common structure but leaves the predicates and their sorts unspecified, since they vary across domains. 
Grammar induction thus reduces to constructing the domain-specific vocabulary of predicates and sorts.

GrammarAgent processes the rules one at a time, each with the associated entity's declaration.
Generating a predicate directly from a rule is unreliable, because a rule often combines multiple requirements and a condition. 
Naming a predicate before this structure is disentangled 
may make it too large and specific to that one rule, or even reverse the rule's meaning
(\eg, turning ``throw unless authorized'' into ``throw if authorized'').
\emph{GrammarAgent} therefore uses two LLM passes. 
It first splits each rule into atoms, which are independently checkable conditions or obligations, 
and the Boolean formulas connecting them (\stepname{Extract Atoms}). 
It then names each atom as a predicate and gives every argument a sort (\stepname{Infer Predicates}). 
A Python tool checks each pass (\stepname{Check Atoms}, \stepname{Check Predicates}), verifying that formulas refer to the correct atoms and that all atoms are translated, and a final tool collects the predicates and sorts to build the grammar (\stepname{Generate Grammar}).

The prompt for \stepname{Extract Atoms} is shown in Listing~\ref{lst:g1:template} (appendix). It defines an atom as one independently falsifiable observable relation 
and gives detailed guidance on what should become an atom 
and how a rule should be split. For example, when one relation can hold while another fails, it directs the LLM to split the two into separate atoms. 
As another example, it requires each atom to be grounded in a concrete observation (\eg, a state transition or an emitted effect). 
The prompt also guards against common mistakes. 
Finally, it requires the LLM to label each atom as an obligation or a condition and to output the Boolean formulas combining them into the rule's logical structure, 
which the last agent later uses to formalize the rule.

The prompt for \stepname{Infer Predicates} is shown in Listing~\ref{lst:g3:template} (appendix). 
It instructs the LLM to translate the extracted atoms into the smallest vocabulary of predicates and sorts. 
It requires a one-to-one mapping in which every atom yields exactly one predicate. Each predicate inherits its atom's classification, becoming an obligation or condition predicate that guides formal rule generation. 
Each predicate and sort also carries a natural-language description.
The prompt further enforces consistency: a parameter name that is shared across atoms and denotes one semantic variable must take the same sort wherever it occurs, and every referenced sort must be declared in the same response. 
Finally, it forbids encoding concrete values or the whole rule into a predicate name, keeping predicate names small and general.

For example, as shown in Figure~\ref{fig:agent-output-grammar} (appendix), 
the highlighted rule in Figure~\ref{fig:overview} yields two predicates and one sort. 
The obligation predicate \texttt{ThrowObligation()} denotes that the function reverts execution, the condition predicate \texttt{AuthorizedCondition(Address, Address)} 
denotes that the first address has authorized the second, and the single sort \texttt{Address} represents an Ethereum address.

\newcolumntype{Y}{>{\centering\arraybackslash}X}
\begin{table*}[t]
\centering
\footnotesize
{
\mycaption{tab:baseline}
{Precision and recall of rule formalization.}
{\textit{(The \Tool{} columns use GPT-OSS-20B; SiO and GiO use GPT-5.6. SymGPT has no Rust results.)}}
\vspace{0.04in}


\setlength{\tabcolsep}{1.15pt}
\renewcommand{\arraystretch}{1.2}

\begin{tabularx}{\textwidth}{|l|*{7}{Y|}Y||*{6}{Y|}}
\hline
\multirow{3}{*}{\textbf{API Doc}}
  & \multicolumn{8}{c||}{\textbf{Comparison with Baselines}}
  & \multicolumn{6}{c|}{\textbf{Comparison with LLM Backbones}} \\
\cline{2-9}\cline{10-15}

  & \multicolumn{2}{c|}{\textbf{GiO}}
  & \multicolumn{2}{c|}{\textbf{SiO}}
  & \multicolumn{2}{c|}{\textbf{SymGPT}}
  & \multicolumn{2}{c||}{\textbf{Doc2Spec}}
  & \multicolumn{2}{c|}{
      \shortstack[c]{\textbf{OSS-120B}}}
  & \multicolumn{2}{c|}{\textbf{DeepSeek-V4}}
  & \multicolumn{2}{c|}{\textbf{GPT-5.6}} \\
\cline{2-15}

  & \textbf{Prec.} & \textbf{Rec.}
  & \textbf{Prec.} & \textbf{Rec.}
  & \textbf{Prec.} & \textbf{Rec.}
  & \textbf{Prec.} & \textbf{Rec.}
  & \textbf{Prec.} & \textbf{Rec.}
  & \textbf{Prec.} & \textbf{Rec.}
  & \textbf{Prec.} & \textbf{Rec.} \\
\hline\hline

ERC20
  & 0.38 & 0.56 & 0.56 & 0.79 & 1.00 & 0.72 & 0.79 & 0.90 & 0.72 & 0.83 & 0.79 & 0.83 & 0.91 & 1.00 \\
ERC721
  & 0.42 & 0.80 & 0.53 & 0.94 & 0.94 & 0.77 & 0.56 & 0.78 & 0.51 & 0.74 & 0.68 & 0.67 & 0.59 & 0.77 \\
ERC1155
  & 0.11 & 0.63 & 0.13 & 0.67 & 1.00 & 0.80 & 0.37 & 0.73 & 0.41 & 0.83 & 0.49 & 0.80 & 0.47 & 0.80 \\
\hline
Rust-1
  & 0.81 & 0.38 & 0.75 & 0.23 & \NA & \NA & 0.80 & 0.78 & 0.74 & 0.77 & 0.71 & 0.72 & 0.82 & 0.91 \\
Rust-2
  & 0.75 & 0.28 & 0.73 & 0.09 & \NA & \NA & 0.82 & 0.77 & 0.81 & 0.77 & 0.78 & 0.75 & 0.85 & 0.89 \\
Rust-3
  & 0.53 & 0.75 & 0.46 & 0.67 & \NA & \NA & 0.61 & 0.75 & 0.60 & 0.70 & 0.60 & 0.73 & 0.76 & 0.92 \\
\hline
\textbf{Average}
  & \textbf{0.45} & \textbf{0.41} & \textbf{0.43} & \textbf{0.30} & \NA & \NA & \textbf{0.73} & \textbf{0.78} & \textbf{0.70} & \textbf{0.77} & \textbf{0.71} & \textbf{0.74} & \textbf{0.78} & \textbf{0.89} \\
\hline
\end{tabularx}}
\vspace{-0.12in}
\end{table*}

\subsection{Stage 4: Formalizing Rules}
\label{sec:formal}

\emph{FormalAgent} formalizes each natural-language rule into one DSL statement, 
using the predicates, sorts, and Boolean formulas extracted from the rule, together with the rule's associated entity and its attributes. 
As specified by the prompt in Listing~\ref{lst:f2:template} (appendix), the statement must follow the structure in Figure~\ref{fig:template}: the LLM does not re-derive the logical structure but fills in the extracted predicates and prefixes the appropriate modality. The LLM further instantiates 
each predicate by replacing the sorts in its signature with concrete arguments. A Python tool built on the Lark parser then checks that the DSL statement conforms to the grammar (\stepname{Check DSL}).
For example, Figure~\ref{fig:agent-output-formal} (appendix) shows the DSL statement that \emph{FormalAgent} 
generates for the highlighted sentence 
in Figure~\ref{fig:overview}, where the sorts of \texttt{AuthorizedCondition()} are instantiated with the concrete parameters \texttt{\_from} and \texttt{sender} of \texttt{transferFrom()}.

\section{Evaluation}
\label{sec:eval}


\boldparagraph{Benchmarks.}
We evaluate \Tool{} using six API documents: three ERC standards and three Rust source files. 
The ERC standards specify how to implement public APIs for smart contracts managing fungible tokens, non-fungible tokens (NFTs), and both. 
They are also studied in SymGPT~\citep{SymGPT}, whose authors identify 90 implementation rules. The three Rust files are from the Rust standard library 
and cover pointer arithmetic and memory allocation. 
Their API comments contain 425 natural-language rules on how these APIs should be used. Together, the benchmarks span two programming languages (Solidity and Rust) and cover both API implementation rules and API usage rules.
Table~\ref{tab:benchmark} (appendix) shows more detailed information.

\boldparagraph{Baselines.}
We compare \Tool{} against three baselines. 
(1) \emph{Spec-in-One} (SiO) asks a commercial LLM (GPT-5.6) to identify and formalize all rules in a single prompt, with no grammar or staged decomposition. 
(2) \emph{Grammar-in-One} (GiO) first summarizes a grammar in one prompt, then uses it to formalize the rules with FormalAgent.
(3) \emph{SymGPT}~\citep{SymGPT} formalizes ERC rules using manually crafted grammars. 
SiO measures the benefit of grammar guidance, 
GiO isolates the benefit of staged grammar induction, 
and SymGPT provides a comparison with hand-written grammars.
The prompts for SiO and GiO are in Section~\ref{sec:app-baseline-prompts} (appendix).

We do not compare with NL2Type~\citep{malik2019nl2type}, TypeWriter~\citep{pradel2020typewriter}, REQ2LTL~\citep{Ma2025req2ltl}, 
or CLOVER~\citep{ryu_divide_2025} because these tools either focus on 
inferring parameter and return types, focus on formalizing system-level rules 
rather than API rules, or support only a limited set of predefined sorts.

\boldparagraph{LLMs.}
Unless otherwise stated, \Tool{} uses GPT-OSS-20B, 
a 20B-parameter open-source model~\citep{openai2025gptoss120bgptoss20bmodel}. 
To study how the backbone affects \Tool{}, we also evaluate it with 
GPT-OSS-120B~\citep{openai2025gptoss120bgptoss20bmodel}, DeepSeek-V4~\citep{deepseek2026v4}, 
and the frontier model GPT-5.6~\citep{openai2026gpt56sol}. 
These cover scalability within a model family, generalizability across families, 
and frontier-level models.

\boldparagraph{Metrics.}
We compare the formalized rules against 
a gold-standard set of natural-language rules. 
For the three ERCs, we reuse the rule set from SymGPT. 
For the Rust documents, we manually construct the gold-standard set from each document. 
Since the tool produces formal rules, at least two authors manually judge whether each one correctly formalizes a gold-standard rule, 
following a shared rubric; the agreement rate between authors is 87\% (Cohen's $\kappa = 0.75$). 
A gold-standard rule is a true positive (TP) if fully matched and a false negative (FN) otherwise, and an unmatched formal rule is a false positive (FP). We then compute precision and recall accordingly. 
The full rubric is in Section~\ref{sec:app-evaluation-rubric} (appendix).

\boldparagraph{Configurations.}
When \Tool{} detects errors in an LLM output with its Python tools, it prompts the LLM to retry, up to three times. The temperature is set to 1 and the reasoning effort is set to medium for all experiments. Each configuration is run \emph{three} times, 
and we report the average.

\subsection{Effectiveness}
\label{sec:baseline}

As shown in Table~\ref{tab:baseline}, 
\Tool{} achieves an average precision of $0.73$, ranging from $0.37$ on ERC1155 to $0.82$ on Rust-2, and an average recall of $0.78$, with every benchmark reaching at least $0.73$. 
We observe no clear precision--recall trade-off and no correlation between document length and performance. 
The three ERCs consist of API implementation rules and the three Rust documents of API usage rules. \Tool{} handles both: recall is comparable across the two, and is slightly higher on the implementation rules.

Compared with GiO and SiO, 
\Tool{} is substantially better on both metrics: it reaches 
$0.73$ precision and $0.78$ recall, versus 
$0.45$/$0.41$ for GiO and 
$0.43$/$0.30$ for SiO. 
This holds even though both baselines use a stronger backbone (GPT-5.6) than \Tool{} (GPT-OSS-20B). Against SiO, 
\Tool{} improves precision on all six benchmarks and recall on all but ERC721; 
against GiO, it improves precision on five and recall on four. 
These gains show the benefit of \emph{grammar-guided formalization} and of \emph{inducing the grammar through staged decomposition} rather than in one step.






Compared with SymGPT, \Tool{} achieves lower precision across all three ERCs, as SymGPT is primarily a bug-detection tool focused on result accuracy. 
However, \Tool{} attains higher recall on ERC20 and ERC721, for two reasons. 
First, SymGPT does not define a grammar for return-value rules, whereas \Tool{} infers one and can thus formalize the rules with it. 
Second, SymGPT avoids designing grammars for underspecified rules 
(\eg, ``throw on any error''), whereas \Tool{} can infer grammars 
for such rules.
Overall, these recall gains over SymGPT show that \emph{\Tool{}'s automatic grammar induction can match or exceed hand-written grammars in many cases.}

Finally, we evaluate \Tool{} with different backbones. Precision ranges from $0.70$ to $0.78$ and recall from $0.74$ to $0.89$.
Consistency across model sizes and families indicates that
\emph{\Tool{}'s stability comes from its staged, grammar-guided
design rather than any single backbone.}
\begin{figure}[t]
  \centering  \includegraphics[width=0.7\linewidth,keepaspectratio]{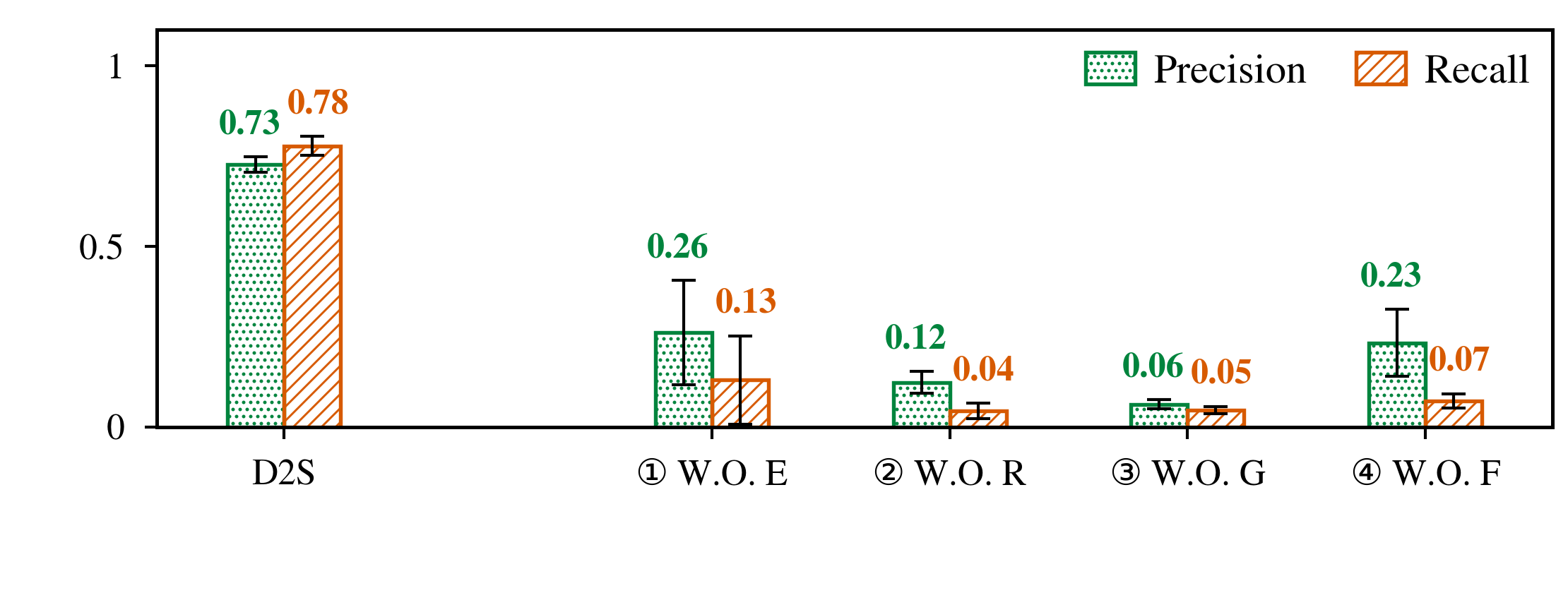}
  \vspace{-0.2in}
  \mycaption{fig:ablation}
 {\Tool{} ablation results.}
    {\textit{(W.O.: without; E: EntityLocalizer; R: NLRuleAgent; G: GrammarAgent; F: FormalAgent. Each ablation replaces one agent with a single prompt.)}}
    
\end{figure}

\subsection{Ablation Analysis}
\label{sec:eval-ablation}

To understand each agent's contribution, we ablate one agent at a time by replacing it with a single-prompt alternative and measuring how \Tool{}'s precision and recall change over the six benchmarks. 
For EntityLocalizer, we give the LLM the full document and ask it to identify all 
entities directly. For NLRuleAgent, we provide the document 
and extracted entities and ask it to identify the rule-stating sentences for each entity. For GrammarAgent, we provide all rules and the DSL template and ask it to summarize the sorts and predicates. 
For FormalAgent, we provide all entities, attributes, rules, and the full grammar in one prompt and ask it to produce the formalized DSL rules.

As shown in Figure~\ref{fig:ablation}, removing any single agent sharply degrades \Tool{}. 
Precision drops to $0.26$, $0.12$, $0.06$, and $0.23$ without EntityLocalizer, NLRuleAgent, GrammarAgent, and FormalAgent respectively, 
and recall to $0.13$, $0.04$, $0.05$, and $0.07$. These results show that all four agents contribute substantially and \emph{validate the effectiveness of \Tool{}'s multi-agent design}.

\subsection{Downstream Verification}
\label{sec:verification}

To show that the rules formalized by \Tool{} are correct and useful, we use existing general-purpose symbolic execution engines (SymGPT's backend
for Solidity smart contracts and KLEE~\citep{klee} for Rust) 
to check real code against the rules and detect real violations.

Our evaluated code set contains 200 Ethereum smart contracts (181 for ERC20, 12 for ERC721, and 7 for ERC1155) and 20 Rust crates from 
crates.io that either implement the APIs or exercise the APIs our specifications cover. 
The formalized rules cannot be checked directly: 
a symbolic execution engine needs to know when a violation occurs, 
and some predicates refer to program states that do not have corresponding program variables. 
We therefore use an LLM to instrument each program with the required ghost variables and assertions before running the engine. 
The full pipeline is described in Appendix~\ref{sec:app-verifier-workflow}.

We compare against nine baselines. 
(1)	\emph{LLM-on-Docs} asks GPT-5.6 to check each pair of a natural-language rule document 
and a source file and judge whether any rule is violated. 
(2)	\emph{LLM-on-Rules} instead gives GPT-5.6 each rule formalized by \Tool{}, 
paired with each source file, and asks the same question, isolating the effect of formalization. 
(3)--(9) SymGPT and six static Rust bug-detection tools (Semgrep, LockBud, Prusti, Rudra, MIRAI, and MirChecker). Detailed baseline information is in Table~\ref{tab:verification-tools} (appendix). We count real code sites where a rule is violated (TP) and false alarms (FP).
As shown in Table 5 (appendix), Doc2Spec with symbolic execution detects 142 real violations with only 21 false alarms, confirming \emph{the formalized rules’ correctness and practical usefulness.}

Compared with our approach, the baselines LoD and LoR report many true positives, and LoR even detects more than \Tool{}. However, they produce hundreds of false alarms (324 and 602, respectively), which demonstrates that our pipeline is essential for precise detection of rule violations.
\Tool{} trails SymGPT (268 TPs, 6 FPs) on the ERC benchmarks, but SymGPT relies on a hand-written grammar, whereas \Tool{} induces the grammar fully automatically. On Rust, \Tool{} detects 24 violations with no false alarms, exceeding all six specialized tools combined (13 TPs, 90 FPs); 
these tools must encode each checked rule through dedicated program analysis and thus scale poorly to new rules.

\section{Conclusion}
This paper presents \Tool{}, a multi-agent framework that automatically formalizes natural-language rules on API implementation and usage. 
Our key insight is to decompose specification generation into ordered, constrained steps: \Tool{} fixes a domain-agnostic logical skeleton as a grammar template, 
induces domain-specific predicates and sorts from the natural-language rules to form a grammar, 
and uses the grammar to guide rule formalization. 

Across six benchmarks spanning Solidity and Rust and covering both API implementation and API usage, 
\Tool{} improves precision by 0.28 and recall by 0.37 over baselines that lack grammar induction or perform it in one unstaged step.
Integrated with downstream symbolic-execution tools, its formalized rules uncover 142 previously unknown rule violations in real-world smart contracts and Rust crates, with only 21 false alarms. These results demonstrate the benefits of grammar induction for LLM-based specification generation. 
We hope \Tool{} serves as a step toward making formal verification accessible directly from natural-language documentation.

\section*{AI Use Statement}

Large language models are components of the proposed Doc2Spec
framework and the experimental evaluation. They are used to
analyze API documentation, extract natural-language rules,
construct domain-specific grammars, and translate rules into
formal specifications. The models, prompts, and validation
procedures are described in the main text and appendices.

During manuscript preparation, we also used generative AI tools
to assist with language editing, literature search, LaTeX
formatting, and the preparation and refinement of figures and
tables.

The authors reviewed and revised the AI-assisted material,
checked technical claims and references against the original
sources, and take full responsibility for the final content
of this work, including its text, methods, results, and artifacts.
\bibliography{ir} 
\bibliographystyle{iclr2027_conference}
\newpage
\appendix
\clearpage
\onecolumn
\raggedbottom
\part*{Appendix}

\section{Agent Outputs}
\label{sec:app-agent-outputs}
\begingroup
\captionsetup{skip=3pt}
\setlength{\intextsep}{6pt plus 2pt minus 2pt}
\definecolor{OutputEntity}{HTML}{E66B96}
\definecolor{OutputRule}{HTML}{E8C442}
\definecolor{OutputGrammar}{HTML}{4EB3C5}
\definecolor{OutputFormal}{HTML}{98C45E}
\tcbset{agentoutput/.style={
  enhanced, sharp corners, boxrule=0.5pt, boxsep=1pt,
  left=7pt, right=7pt, top=3pt, bottom=3pt,
  before skip=0pt, after skip=0pt,
  fonttitle=\sffamily\bfseries\small, coltitle=black,
  fontupper=\fontsize{8.5}{9.5}\selectfont,
  toptitle=2pt, bottomtitle=2pt,
}}
\lstdefinestyle{agentoutputcode}{
  basicstyle=\ttfamily\fontsize{8}{9.5}\selectfont,
  columns=fullflexible, keepspaces=true, showstringspaces=false,
  breaklines=true, breakatwhitespace=true,
  numbers=none, frame=none, backgroundcolor={},
  aboveskip=2pt, belowskip=0pt,
  xleftmargin=0pt, xrightmargin=0pt,
}

\begin{figure}[!htbp]
\centering
\begin{tcolorbox}[agentoutput,
  colback=OutputEntity!6, colframe=OutputEntity!80!black,
  colbacktitle=OutputEntity!75,
  title={EntityLocalizer Output Example}]
\textbf{Target declaration}
\begin{lstlisting}[style=agentoutputcode]
function transferFrom(address _from, address _to, uint256 _value)
  public returns (bool success)
\end{lstlisting}
\smallskip
\textbf{Attributes}\quad
\texttt{kind: function; name: transferFrom; visibility: public}
\par
\textbf{Parameters (ordered)}\quad
\texttt{\_from: address, \_to: address, \_value: uint256}
\par
\textbf{Return}\quad\texttt{success: bool}
\qquad\textbf{Hierarchy}\quad\texttt{[]}
\par
\textbf{Attached NL}
\par
Transfers \texttt{\_value} amount of tokens from address \texttt{\_from} to address
\texttt{\_to}, and MUST fire the Transfer event.
\par\smallskip
The transferFrom method is used for a withdraw workflow, allowing contracts to
transfer tokens on your behalf. This can be used for example to allow a contract
to transfer tokens on your behalf and/or to charge fees in sub-currencies.
The function SHOULD throw unless the \texttt{\_from} account has deliberately
authorized the sender of the message via some mechanism.
\par\smallskip
Note Transfers of 0 values MUST be treated as normal transfers and fire the
Transfer event.
\end{tcolorbox}
\vspace{0.06in}
\caption{EntityLocalizer output for the highlighted example in Figure~\ref{fig:overview}.}
\label{fig:agent-output-entity}
\end{figure}

\begin{figure}[!htbp]
\centering
\begin{tcolorbox}[agentoutput,
  colback=OutputRule!9, colframe=OutputRule!70!black,
  colbacktitle=OutputRule!85,
  title={NLRuleAgent Output Example}]
The function SHOULD throw unless the \texttt{\_from} account has deliberately
authorized the sender of the message via some mechanism.
\par\smallskip
\textbf{Candidate}\quad\texttt{candidate-0003}
\par\smallskip
\textbf{Score Clause}\quad
Evaluation 1: $r_1=+1$ (\texttt{true}), $c_1=0.8$;
\quad Evaluation 2: $r_2=+1$ (\texttt{true}), $c_2=0.8$.
\par\smallskip
\textbf{Filter Clause}\quad
$\displaystyle\text{score}(s)
\approx 0.8320 > 0.5$.
\par\smallskip
\textbf{Result}\quad Retain the clause as a rule.
\end{tcolorbox}
\vspace{0.06in}
\caption{NLRuleAgent output for the highlighted example in Figure~\ref{fig:overview}.}
\label{fig:agent-output-rule}
\end{figure}

\begin{figure}[!htbp]
\centering
\begin{tcolorbox}[agentoutput,
  colback=OutputGrammar!6, colframe=OutputGrammar!75!black,
  colbacktitle=OutputGrammar!75,
  title={GrammarAgent Output Example}]
\textbf{Sort}\quad\texttt{Address}: Ethereum address
\par\smallskip
\textbf{Semantic atoms $\rightarrow$ predicate signatures}
\begin{lstlisting}[style=agentoutputcode]
O_throw      -> ThrowObligation()
C_authorized -> AuthorizedCondition(_from: Address, sender: Address)
\end{lstlisting}
\smallskip
\textbf{Outcome formula}\quad\texttt{O\_throw}
\qquad\textbf{Condition formula}\quad\texttt{not C\_authorized}
\par\smallskip
\end{tcolorbox}
\vspace{0.06in}
\caption{GrammarAgent output for the highlighted example in Figure~\ref{fig:overview}.}
\label{fig:agent-output-grammar}
\end{figure}

\begin{figure}[!htbp]
\centering
\begin{tcolorbox}[agentoutput,
  colback=OutputFormal!8, colframe=OutputFormal!75!black,
  colbacktitle=OutputFormal!80,
  title={FormalAgent Output Example}]
\begin{lstlisting}[style=agentoutputcode]
SHOULD ThrowObligation() if not AuthorizedCondition(_from, sender);
\end{lstlisting}
\end{tcolorbox}
\vspace{0.06in}
\caption{FormalAgent output for the highlighted example in Figure~\ref{fig:overview}.}
\label{fig:agent-output-formal}
\end{figure}
\endgroup

\clearpage

\par\noindent
\begin{minipage}{\linewidth}
\section{Benchmarks}
\label{sec:app-api-documents}
\begingroup
  \captionsetup{type=table}
  \centering
  \footnotesize
  \mycaption{tab:benchmark}
    {Benchmark Information.}{}
  \vspace{0.15in}
  \setlength{\tabcolsep}{1.15pt}
  \renewcommand{\arraystretch}{1.2}
  \begin{tabular}{|l|c|c|c|}
\hline
\textbf{API Doc} & \textbf{Lines} & \textbf{Entities} & \textbf{Rules} \\
\hline\hline
ERC20~\citep{erc20}   & 88    & 11 & 21  \\
ERC721~\citep{erc721}  & 436   & 19 & 41  \\
ERC1155~\citep{erc1155} & 710   & 12 & 28  \\
\hline
Rust-1~\citep{rust_const_ptr}  & 1,741 & 47 & 168 \\
Rust-2~\citep{rust_mut_ptr}  & 2,157 & 62 & 226 \\
Rust-3~\citep{rust_alloc_global}  & 277   & 5  & 31  \\
\hline\hline
\textbf{Total} & \textbf{5,409} & \textbf{156} & \textbf{515} \\
\hline
\end{tabular}

  \par
\endgroup

\end{minipage}
\par

\section{Rule Evaluation Rubric}
\label{sec:app-evaluation-rubric}

Each generated DSL rule is paired with an extracted natural-language (NL)
clause and a target entity. Reviewers use the NL clause to identify candidate
gold-standard requirements, then check entity grounding and the actual DSL
semantics. The procedure
below uses the gold-to-generated direction when naming match cardinalities.

\begin{enumerate}[leftmargin=*, itemsep=5pt]
    \item \textbf{Find candidate matches using the extracted NL clauses.}
    Compare each extracted clause with the gold-standard NL rules. Identify
    candidates that describe the same requirement, or complementary parts of
    that requirement. Compare meaning rather than wording: retain the required
    behavior, conditions, exceptions, and modality. An extracted clause with no
    corresponding gold requirement cannot yield an accepted match.

    \item \textbf{Check the associated entity.}
    For each candidate, verify the declaration, declaration attributes, and
    lexical hierarchy against the gold target. Names alone are insufficient:
    distinguish overloads, enclosing scopes, parameter and return roles, and
    receiver information where relevant. Reject candidates bound to the wrong
    entity. Entity correctness is a prerequisite, not a separate counting unit.

    \item \textbf{Determine the candidate match.}
    After entity matching, distinguish one gold rule paired with one generated
    rule (one-to-one), one gold rule split across several generated rules
    (one-to-many), and several gold rules merged into one generated rule
    (many-to-one). Use only rules needed to express the candidate requirement;
    do not combine unrelated targets or add duplicates to a group.

    \item \textbf{Validate the DSL for the identified case.}
    Missing, malformed, or placeholder DSL cannot participate in an accepted
    match. Interpret parseable DSL using its predicate definitions. In every
    case, check predicate meaning; argument roles, bindings, sorts, constants,
    and negation; the presence and composition of conditions; Boolean operators
    and scope; modality; and all consequences, exceptions, and required ordering.
    Reject omissions, contradictions, unsupported additions, strengthening, or
    weakening. Apply these checks as follows:
    \begin{description}[leftmargin=1.4em, style=nextline, itemsep=3pt]
        \item[One-to-one.]
        The generated DSL must express the complete meaning of the gold rule.
        Matching the attached NL clause or a predicate name is insufficient.
        \item[One-to-many.]
        The smallest necessary set of generated DSL rules must jointly express
        the complete gold requirement. Every required consequence must retain
        its own condition and modality; splitting a rule must not make a
        conditional obligation unconditional. Partial coverage is not a TP.
        \item[Many-to-one.]
        The single generated DSL rule must fully express each gold requirement
        in the group, preserving each condition--consequence association and
        modality. Do not credit several gold rules merely because their topics
        appear in one extracted NL clause. If only a subset is covered, accept
        that subset only when the complete DSL is equivalent to it and contains
        no unsupported obligations.
    \end{description}
    Repartition a more complex candidate into these cases only when the
    resulting groups independently satisfy all checks; do not accept an
    otherwise unresolved many-to-many group.

    \item \textbf{Assign TP, FP, and FN.}
    Assign each gold rule and each generated rule to at most one accepted group.
    Count each covered gold rule as one TP and each uncovered gold rule as one
    FN. Count each generated rule outside accepted groups as one FP, including
    unnecessary duplicates, invalid DSL, incorrect targets, and predictions
    without a gold requirement. Thus, a valid one-to-many group contributes one
    TP; a valid many-to-one group covering $k$ gold rules contributes $k$ TPs.
    A rejected candidate leaves its gold rules as FNs unless another valid group
    covers them, and its generated rules as FPs unless accepted elsewhere.
\end{enumerate}


\section{Integration with symbolic execution engines}
\label{sec:app-verifier-workflow}

The pipeline takes the rules formalized by \Tool{} and the program code as input. 
It contains two agents that process each pair of an input program and a formalized rule in turn. The first agent inserts the ghost variables that represent the program states required by the rule, 
together with the updates to these variables along the program. 
The second agent instruments assertions at the appropriate program sites, 
so that an assertion fails exactly when the rule is violated. The symbolic execution engine, either SymGPT's backend or KLEE, then searches for an input that triggers each inserted assertion. 
If such an input is found, a rule violation is detected; 
otherwise, the program is verified to satisfy the rule.

\begin{figure}[ht]
  \centering
  \includegraphics[width=\textwidth]{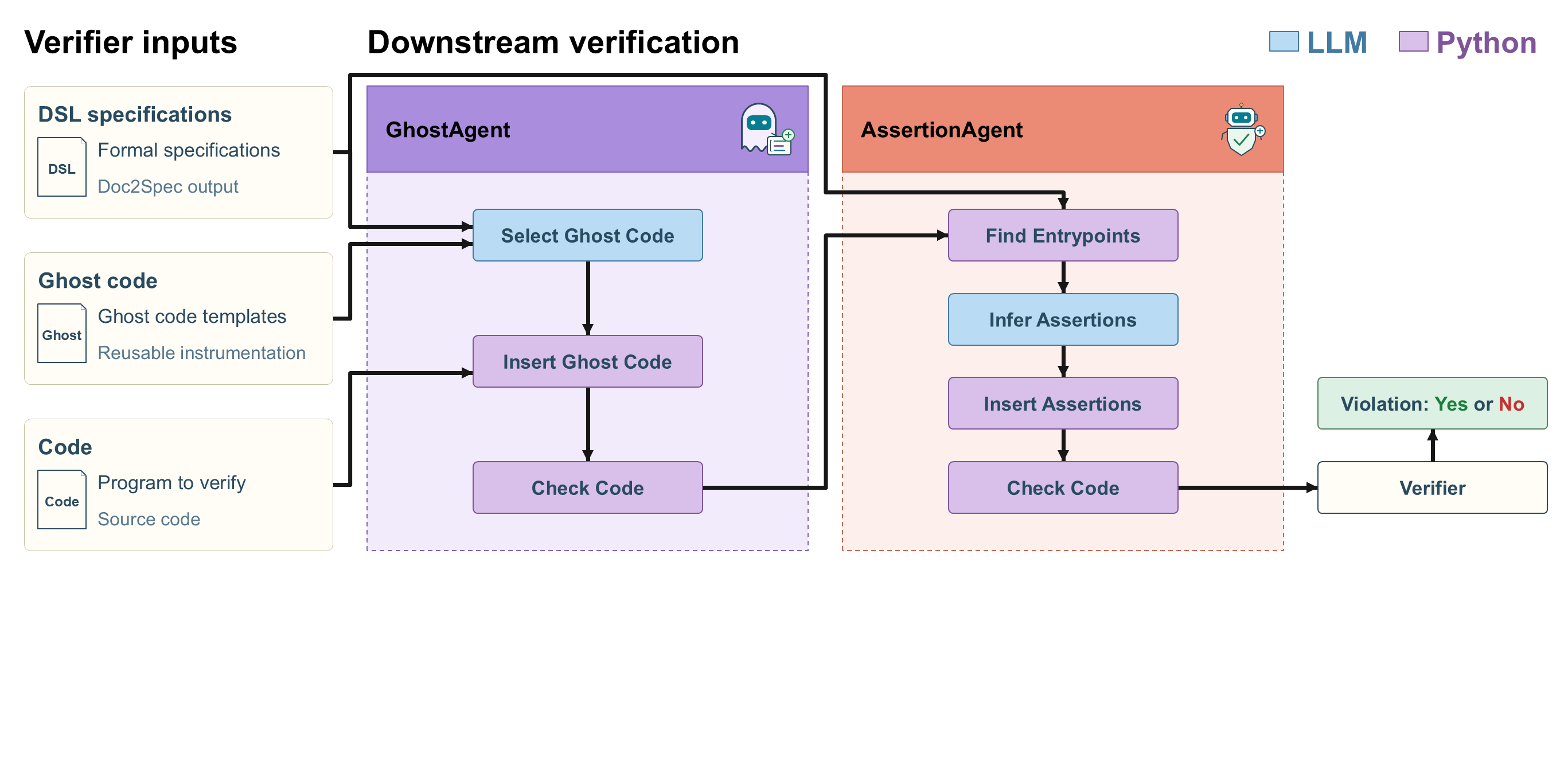}
  \vspace{-0.75in}
  \mycaption{fig:verifier-architecture}{Integration with downstream symbolic execution engines.}{}
\end{figure}

\par\noindent
\begin{minipage}{\linewidth}
\section{Baselines for rule violation detection}
\label{sec:app-verification-tools}
\begingroup
\captionsetup{type=table}
\centering
\footnotesize
\mycaption{tab:verification-tools}
  {Baselines used in the downstream rule-violation detection.}
  {\textit{(Capability summarizes their primary
analysis focus. Language indicates the programming language analyzed by each tool. 
Stars reports the number of GitHub stars. A gray cell denotes an unavailable GitHub star count.)}}
\vspace{0.15in}
\setlength{\tabcolsep}{6pt}
\renewcommand{\arraystretch}{1.2}
\begin{tabular}{|l|c|c|c|}
\hline
\textbf{Tool} & \textbf{Capability} & \textbf{Language} & \textbf{Stars} \\
\hline\hline
LLM-on-Docs & Rule compliance & Solidity \& Rust & \NA \\
LLM-on-Rules & Rule compliance & Solidity \& Rust & \NA \\
\hline
\href{https://github.com/symbolic-gpt/symgpt}{SymGPT}~\citep{SymGPT} & ERC compliance & Solidity & 8 \\
\hline
\href{https://github.com/semgrep/semgrep}{Semgrep}~\citep{semgrep} & API misuse & Rust & 16,725 \\
\href{https://github.com/BurtonQin/lockbud}{LockBud}~\citep{lockbud} & Concurrency & Rust & 591 \\
\href{https://github.com/prusti/prusti}{Prusti}~\citep{prusti} & Correctness & Rust & 1,811 \\
\href{https://github.com/sslab-gatech/Rudra}{Rudra}~\citep{rudra} & Memory safety & Rust & 1,367 \\
\href{https://github.com/endorlabs/MIRAI}{MIRAI}~\citep{mirai} & Correctness & Rust & 229 \\
\href{https://github.com/lizhuohua/rust-mir-checker}{MirChecker}~\citep{mirchecker} & Memory safety & Rust & 165 \\
\hline
\end{tabular}

\par
\endgroup
\end{minipage}
\par

\par\noindent
\begin{minipage}{\linewidth}
\section{Rust Crate Benchmarks}
\label{sec:app-rust-crates}
\begingroup
\captionsetup{type=table}
\centering
\footnotesize
\mycaption{tab:rust-crate-apis}
  {Information about the Rust crates.}
  {\textit{(Stars reports the number of GitHub stars.
    A gray cell denotes an unavailable GitHub star count.)}}
\vspace{0.15in}
\setlength{\tabcolsep}{6pt}
\renewcommand{\arraystretch}{1.2}
\begin{tabular}{|l|c|}
\hline
\mbox{\textbf{Crate (version)}} & \mbox{\textbf{Stars}}  \\
\hline\hline
\href{https://crates.io/crates/aegis/0.9.12}{aegis (0.9.12)} & \href{https://github.com/jedisct1/rust-aegis}{54}  \\
\href{https://crates.io/crates/cyberex/0.3.12}{cyberex (0.3.12)} & \NA  \\
\href{https://crates.io/crates/falcon_capstone/0.5.2}{falcon\_capstone (0.5.2)} & \href{https://github.com/falconre/falcon_capstone}{4}  \\
\href{https://crates.io/crates/fast-able/1.11.17}{fast-able (1.11.17)} & \NA \\
\href{https://crates.io/crates/fast-float/0.2.0}{fast-float (0.2.0)} & \href{https://github.com/aldanor/fast-float-rust}{292}  \\
\href{https://crates.io/crates/grubbnet/0.1.7}{grubbnet (0.1.7)} & \href{https://github.com/Dooskington/grubbnet}{77}  \\
\href{https://crates.io/crates/halo2curves-axiom/0.7.0}{halo2curves-axiom (0.7.0)} & \href{https://github.com/axiom-crypto/halo2curves}{7}  \\
\href{https://crates.io/crates/hugepage-rs/0.1.0}{hugepage-rs (0.1.0)} & \href{https://github.com/cppcoffee/hugepage-rs}{13}  \\
\href{https://crates.io/crates/i_tree/0.9.0}{i\_tree (0.9.0)} & \href{https://github.com/iShape-Rust/iTree}{1}  \\
\href{https://crates.io/crates/i_triangle/0.26.0}{i\_triangle (0.26.0)} & \href{https://github.com/iShape-Rust/iTriangle}{160}  \\
\href{https://crates.io/crates/jh/0.1.0}{jh (0.1.0)} & \href{https://github.com/RustCrypto/hashes}{2,266} \\
\href{https://crates.io/crates/jomini/0.27.0}{jomini (0.27.0)} & \href{https://github.com/rakaly/jomini}{90}  \\
\href{https://crates.io/crates/lurk-cli/0.3.8}{lurk-cli (0.3.8)} & \href{https://github.com/JakWai01/lurk}{1,154}  \\
\href{https://crates.io/crates/meshopt/0.3.0}{meshopt (0.3.0)} & \href{https://github.com/gwihlidal/meshopt-rs}{223} \\
\href{https://crates.io/crates/ptr_hash/0.1.1}{ptr\_hash (0.1.1)} & \href{https://github.com/RagnarGrootKoerkamp/PtrHash}{103}  \\
\href{https://crates.io/crates/qwt/0.2.0}{qwt (0.2.0)} & \href{https://github.com/rossanoventurini/qwt}{57} \\
\href{https://crates.io/crates/revm-interpreter/10.0.3}{revm-interpreter (10.0.3)} & \href{https://github.com/bluealloy/revm}{2,243}  \\
\href{https://crates.io/crates/roqoqo-quest/0.14.4}{roqoqo-quest (0.14.4)} & \href{https://github.com/HQSquantumsimulations/qoqo-quest}{9}  \\
\href{https://crates.io/crates/splay_tree/0.3.1}{splay\_tree (0.3.1)} & \href{https://github.com/sile/splay_tree}{19}  \\
\href{https://crates.io/crates/stackvector/1.1.1}{stackvector (1.1.1)} & \href{https://github.com/Alexhuszagh/rust-stackvector}{3}  \\
\hline
\end{tabular}

\par
\endgroup
\end{minipage}
\par

\par\noindent
\begin{minipage}{\linewidth}

\section{Downstream Rule-Violation Detection}
\label{sec:app-verification-results}
\centering
\begingroup
\footnotesize
\captionsetup{type=table}
\mycaption{tab:verification}
{Violation detection results.}
{\textit{(LoD denotes LLM-on-Docs, LoR denotes LLM-on-Rules, Semg denotes Semgrep, LockB denotes LockBud, MirC denotes MirChecker, and 
D2S denotes \Tool{}.
$x_{y}$ denotes x true violations and y false alarms.
A gray cell denotes an unsupported programming language.)}}
\vspace{0.15in}
\setlength{\tabcolsep}{1.15pt}
\renewcommand{\arraystretch}{1.2}
\begin{tabularx}{\linewidth}{|>{\raggedright\arraybackslash}X|*{10}{>{\centering\arraybackslash}X|}}
\hline
 & \textbf{LoD} & \textbf{LoR} & \textbf{SymGPT} & \textbf{Semg} & \textbf{LockB} & \textbf{Prusti} & \textbf{Rudra} & \textbf{MIRAI} & \textbf{MirC} & \textbf{D2S}\\
\hline\hline
\textbf{ERC}   & $60_{79}$ & $222_{483}$ & $268_{6}$ & \NA & \NA & \NA & \NA & \NA & \NA  & $118_{21}$ \\
\hline
\textbf{Rust}    & $24_{245}$     & $18_{119}$   & \NA     & $0_{12}$ & $0_{0}$ & $0_{0}$ & $6_{18}$ & $1_{54}$ & $6_{6}$  & $24_{0}$ \\
\hline
\textbf{Total}  & $\mathbf{84_{324}}$ & $\mathbf{240_{602}}$ & $\mathbf{268_{6}}$ & $\mathbf{0_{12}}$ & $\mathbf{0_{0}}$ & $\mathbf{0_{0}}$ & $\mathbf{6_{18}}$ & $\mathbf{1_{54}}$ & $\mathbf{6_{6}}$ & $\mathbf{142_{21}}$ \\
\hline
\end{tabularx}
\endgroup

\end{minipage}
\par
\clearpage

\section{Prompt Templates and Examples}
\label{sec:app-agent-prompts}

\Needspace{16\baselineskip}
\subsection{EntityLocalizer}
\label{app:entity-localizer}

\begingroup
\PromptSetPlaceholderDelimiters{<<}{>>}

\Needspace{14\baselineskip}
\subsubsection{Infer Scope Regex}

\PromptTemplate[
  title={\paHeading{Prompt template}{Infer Scope Regex}},
  caption={The prompt template of \stepname{Infer Scope Regex}.},
  caption note={\textit{(For parameters involved in the template,
    \PromptParam{DOCUMENT_CHUNK} represents the document chunk to analyze.)}},
  label={lst:e2:template},
]
  {Infer Scope Regex}
  {appendices/prompt_artifacts/01_entity_scope_discovery.template.txt}

\PromptParameter[
  title={\paHeading{Input}{Document chunk}},
  caption={Example for parameter \PromptParam{DOCUMENT_CHUNK} in Listing~\ref{lst:e2:template}.},
  label={lst:e2:param-document},
]
  {DOCUMENT_CHUNK}
  {appendices/prompt_artifacts/01_entity_scope_discovery.parameter_document_chunk.txt}

\LLMResponse[
  title={\paHeading{LLM response}{Infer Scope Regex}},
  caption={Example LLM response for \stepname{Infer Scope Regex} using the prompt template in Listing~\ref{lst:e2:template}.},
  label={lst:e2:response},
]
  {Infer Scope Regex}
  {appendices/prompt_artifacts/01_entity_scope_discovery.response.json}

The ERC-20 example contains no enclosing lexical scope. A supplementary ERC-721 example
with a non-empty scope result appears in
Section~\ref{sec:app-entity-localizer-supplement}.

\Needspace{14\baselineskip}
\subsubsection{Infer Entity Regex}

\PromptTemplate[
  title={\paHeading{Prompt template}{Infer Entity Regex}},
  caption={The prompt template of \stepname{Infer Entity Regex}.},
  caption note={\textit{(For parameters involved in the template,
    \PromptParam{EXISTING_PATTERNS_JSON} represents the scope patterns discovered by \stepname{Infer Scope Regex},
    and \PromptParam{DOCUMENT_CHUNK} represents the document chunk to analyze.)}},
  label={lst:e3:template},
]
  {Infer Entity Regex}
  {appendices/prompt_artifacts/02_entity_entity_discovery.template.txt}

\PromptParameter[
  title={\paHeading{Input}{Existing patterns}},
  caption={Example for parameter \PromptParam{EXISTING_PATTERNS_JSON} in Listing~\ref{lst:e3:template}.},
  label={lst:e3:param-patterns},
]
  {EXISTING_PATTERNS_JSON}
  {appendices/prompt_artifacts/02_entity_entity_discovery.parameter.existing_patterns_json.txt}

\PromptParameter[
  title={\paHeading{Input}{Document chunk}},
  caption={Example for parameter \PromptParam{DOCUMENT_CHUNK} in Listing~\ref{lst:e3:template}.},
  label={lst:e3:param-document},
]
  {DOCUMENT_CHUNK}
  {appendices/prompt_artifacts/02_entity_entity_discovery.parameter_document_chunk.txt}

\LLMResponse[
  title={\paHeading{LLM response}{Infer Entity Regex}},
  caption={Example LLM response for \stepname{Infer Entity Regex} using the prompt template in Listing~\ref{lst:e3:template}.},
  label={lst:e3:response},
]
  {Infer Entity Regex}
  {appendices/prompt_artifacts/02_entity_entity_discovery.response.json}

\Needspace{14\baselineskip}
\subsubsection{Prune Regex}

\PromptTemplate[
  title={\paHeading{Prompt template}{Prune Regex}},
  caption={The prompt template of \stepname{Prune Regex}.},
  caption note={\textit{(For parameters involved in the template,
    \PromptParam{FULL_DOCUMENT_MATCH_EVIDENCE_JSON} represents candidate patterns and all their matches in the document.)}},
  label={lst:e5:template},
]
  {Prune Regex}
  {appendices/prompt_artifacts/03_entity_pattern_validation.template.txt}

\PromptParameter[
  title={\paHeading{Input}{Match evidence}},
  caption={Example for parameter \PromptParam{FULL_DOCUMENT_MATCH_EVIDENCE_JSON} in Listing~\ref{lst:e5:template}.},
  label={lst:e5:param-evidence},
]
  {FULL_DOCUMENT_MATCH_EVIDENCE_JSON}
  {appendices/prompt_artifacts/03_entity_pattern_validation.parameter.full_document_match_evidence_json.txt}

\LLMResponse[
  title={\paHeading{LLM response}{Prune Regex}},
  caption={Example LLM response for \stepname{Prune Regex} using the prompt template in Listing~\ref{lst:e5:template}.},
  label={lst:e5:response},
]
  {Prune Regex}
  {appendices/prompt_artifacts/03_entity_pattern_validation.output.json}

\Needspace{14\baselineskip}
\subsubsection{Extract Attributes}
\label{sec:example-attribute-value}

\PromptTemplate[
  title={\paHeading{Prompt template}{Extract Attributes}},
  caption={The prompt template of \stepname{Extract Attributes}.},
  caption note={\textit{(For parameters involved in the template,
    \PromptParam{CANDIDATE_RECORDS_JSON} represents candidate declarations with line numbers, attached documentation, and enclosing-scope evidence.)}},
  label={lst:e7:template},
]
  {Extract Attributes}
  {appendices/prompt_artifacts/04_entity_candidate_classification.template.txt}

\PromptParameter[
  title={\paHeading{Input}{Candidate records}},
  caption={Example for parameter \PromptParam{CANDIDATE_RECORDS_JSON} in Listing~\ref{lst:e7:template}.},
  label={lst:e7:param-candidates},
]
  {CANDIDATE_RECORDS_JSON}
  {appendices/prompt_artifacts/04_entity_candidate_classification.parameter.candidate_records_json.txt}

\LLMResponse[
  title={\paHeading{LLM response}{Extract Attributes}},
  caption={Example LLM response for \stepname{Extract Attributes} using the prompt template in Listing~\ref{lst:e7:template}.},
  label={lst:e7:response},
]
  {Extract Attributes}
  {appendices/prompt_artifacts/04_entity_candidate_classification.output.json}

\endgroup

\Needspace{16\baselineskip}
\subsection{NLRuleAgent}
\label{sec:app-nlrule-agent}

\begingroup
\PromptSetPlaceholderDelimiters{<<}{>>}

\Needspace{14\baselineskip}
\subsubsection{Extract Clause}

\PromptTemplate[
  title={\paHeading{Prompt template}{Extract Clause}},
  caption={The prompt template of \stepname{Extract Clause}.},
  caption note={\textit{(For parameters involved in the template,
    \PromptParam{TARGET_DECLARATION} represents the associated entity declaration,
    and \PromptParam{ATTACHED_DOCUMENTATION} represents the natural-language description attached to the entity.)}},
  label={lst:n1:template},
]
  {Extract Clause}
  {appendices/prompt_artifacts/05_rule_clause_extraction.template.txt}

\PromptParameter[
  title={\paHeading{Input}{Target declaration}},
  caption={Example for parameter \PromptParam{TARGET_DECLARATION} in Listing~\ref{lst:n1:template}.},
  label={lst:n1:param-target},
]
  {TARGET_DECLARATION}
  {appendices/prompt_artifacts/05_rule_clause_extraction.parameter.target_declaration.txt}

\PromptParameter[
  title={\paHeading{Input}{Attached documentation}},
  caption={Example for parameter \PromptParam{ATTACHED_DOCUMENTATION} in Listing~\ref{lst:n1:template}.},
  label={lst:n1:param-document},
]
  {ATTACHED_DOCUMENTATION}
  {appendices/prompt_artifacts/05_rule_clause_extraction.parameter.attached_documentation.txt}

\LLMResponse[
  title={\paHeading{LLM response}{Extract Clause}},
  caption={Example LLM response for \stepname{Extract Clause} using the prompt template in Listing~\ref{lst:n1:template}.},
  label={lst:n1:response},
]
  {Extract Clause}
  {appendices/prompt_artifacts/05_rule_clause_extraction.output.json}

\Needspace{14\baselineskip}
\subsubsection{Score Clause}

\PromptTemplate[
  title={\paHeading{Prompt template}{Score Clause}},
  caption={The prompt template of \stepname{Score Clause}.},
  caption note={\textit{(For parameters involved in the template,
    \PromptParam{TARGET_DECLARATION} represents the associated entity declaration,
    \PromptParam{ATTACHED_DOCUMENTATION} represents the natural-language description attached to the entity,
    and \PromptParam{RULE_CANDIDATES_JSON} represents candidate clauses and their identifiers.)}},
  label={lst:n3:template},
]
  {Score Clause}
  {appendices/prompt_artifacts/06_rule_candidate_scoring.template.txt}

\PromptParameter[
  title={\paHeading{Input}{Target declaration}},
  caption={Example for parameter \PromptParam{TARGET_DECLARATION} in Listing~\ref{lst:n3:template}.},
  label={lst:n3:param-target},
]
  {TARGET_DECLARATION}
  {appendices/prompt_artifacts/06_rule_candidate_scoring.parameter.target_declaration.txt}

\PromptParameter[
  title={\paHeading{Input}{Attached documentation}},
  caption={Example for parameter \PromptParam{ATTACHED_DOCUMENTATION} in Listing~\ref{lst:n3:template}.},
  label={lst:n3:param-document},
]
  {ATTACHED_DOCUMENTATION}
  {appendices/prompt_artifacts/06_rule_candidate_scoring.parameter.attached_documentation.txt}

\PromptParameter[
  title={\paHeading{Input}{Candidate clauses}},
  caption={Example for parameter \PromptParam{RULE_CANDIDATES_JSON} in Listing~\ref{lst:n3:template}.},
  label={lst:n3:param-candidates},
]
  {RULE_CANDIDATES_JSON}
  {appendices/prompt_artifacts/06_rule_candidate_scoring.parameter.rule_candidates_json.txt}

\LLMResponse[
  title={\paHeading{LLM response}{Score Clause}},
  caption={Example LLM response for \stepname{Score Clause} using the prompt template in Listing~\ref{lst:n3:template}.},
  caption note={},
  label={lst:n3:response},
]
  {Score Clause}
  {appendices/prompt_artifacts/06_rule_candidate_scoring.output.json}

\endgroup

\Needspace{16\baselineskip}
\subsection{GrammarAgent}
\label{sec:app-grammar-agent}

\begingroup
\PromptSetPlaceholderDelimiters{<<}{>>}

\Needspace{14\baselineskip}
\subsubsection{Extract Atoms}
\label{sec:ga-pred}

\PromptTemplate[
  title={\paHeading{Prompt template}{Extract Atoms}},
  caption={The prompt template of \stepname{Extract Atoms}.},
  caption note={\textit{(For parameters involved in the template,
    \PromptParam{RULE_ID} represents the rule identifier,
    \PromptParam{RULE} represents the original natural-language rule,
    and \PromptParam{TARGET_DECLARATION} represents the associated entity declaration.)}},
  label={lst:g1:template},
]
  {Extract Atoms}
  {appendices/prompt_artifacts/07_predicate_semantic_analysis.template.txt}

\PromptParameter[
  title={\paHeading{Input}{Rule ID}},
  caption={Example for parameter \PromptParam{RULE_ID} in Listing~\ref{lst:g1:template}.},
  label={lst:g1:param-rule-id},
]
  {RULE_ID}
  {appendices/prompt_artifacts/07_predicate_semantic_analysis.parameter.rule_id.txt}

\PromptParameter[
  title={\paHeading{Input}{Rule}},
  caption={Example for parameter \PromptParam{RULE} in Listing~\ref{lst:g1:template}.},
  label={lst:g1:param-clause},
]
  {RULE}
  {appendices/prompt_artifacts/07_predicate_semantic_analysis.parameter.rule.txt}

\PromptParameter[
  title={\paHeading{Input}{Target declaration}},
  caption={Example for parameter \PromptParam{TARGET_DECLARATION} in Listing~\ref{lst:g1:template}.},
  label={lst:g1:param-target},
]
  {TARGET_DECLARATION}
  {appendices/prompt_artifacts/07_predicate_semantic_analysis.parameter.target_declaration.txt}

\LLMResponse[
  title={\paHeading{LLM response}{Extract Atoms}},
  caption={Example LLM response for \stepname{Extract Atoms} using the prompt template in Listing~\ref{lst:g1:template}.},
  label={lst:g1:response},
]
  {Extract Atoms}
  {appendices/prompt_artifacts/07_predicate_semantic_analysis.output.json}

\Needspace{14\baselineskip}
\subsubsection{Infer Predicates}

Listing~\ref{lst:g3:template} shows the prompt template for translating semantic atoms into predicates and sorts.

\PromptTemplate[
  title={\paHeading{Prompt template}{Infer Predicates}},
  caption={The prompt template of \stepname{Infer Predicates}.},
  caption note={\textit{(For parameters involved in the template,
    \PromptParam{RULE_ID} represents the rule identifier,
    \PromptParam{RULE} represents the original natural-language rule,
    \PromptParam{TARGET_DECLARATION} represents the associated entity declaration,
    and \PromptParam{SEMANTIC_ANALYSIS_JSON} represents the validated semantic atoms and Boolean formulas.)}},
  label={lst:g3:template},
]
  {Infer Predicates}
  {appendices/prompt_artifacts/08_predicate_vocabulary_synthesis.template.txt}

\PromptParameter[
  title={\paHeading{Input}{Rule ID}},
  caption={Example for parameter \PromptParam{RULE_ID} in Listing~\ref{lst:g3:template}.},
  label={lst:g3:param-rule-id},
]
  {RULE_ID}
  {appendices/prompt_artifacts/08_predicate_vocabulary_synthesis.parameter.rule_id.txt}

\PromptParameter[
  title={\paHeading{Input}{Rule}},
  caption={Example for parameter \PromptParam{RULE} in Listing~\ref{lst:g3:template}.},
  label={lst:g3:param-clause},
]
  {RULE}
  {appendices/prompt_artifacts/08_predicate_vocabulary_synthesis.parameter.rule.txt}

\PromptParameter[
  title={\paHeading{Input}{Target declaration}},
  caption={Example for parameter \PromptParam{TARGET_DECLARATION} in Listing~\ref{lst:g3:template}.},
  label={lst:g3:param-target},
]
  {TARGET_DECLARATION}
  {appendices/prompt_artifacts/08_predicate_vocabulary_synthesis.parameter.target_declaration.txt}

\PromptParameter[
  title={\paHeading{Input}{Semantic analysis}},
  caption={Example for parameter \PromptParam{SEMANTIC_ANALYSIS_JSON} in Listing~\ref{lst:g3:template}.},
  label={lst:g3:param-analysis},
]
  {SEMANTIC_ANALYSIS_JSON}
  {appendices/prompt_artifacts/08_predicate_vocabulary_synthesis.parameter.semantic_analysis_json.txt}

\LLMResponse[
  title={\paHeading{LLM response}{Infer Predicates}},
  caption={Example LLM response for \stepname{Infer Predicates} using the prompt template in Listing~\ref{lst:g3:template}.},
  label={lst:g3:response},
]
  {Infer Predicates}
  {appendices/prompt_artifacts/08_predicate_vocabulary_synthesis.output.json}

\endgroup

\Needspace{16\baselineskip}
\subsection{FormalAgent}
\label{sec:app-formal-agent}

\begingroup
\PromptSetPlaceholderDelimiters{<<}{>>}

\Needspace{14\baselineskip}
\subsubsection{Infer DSL}

\PromptTemplate[
  title={\paHeading{Prompt template}{Infer DSL}},
  caption={The prompt template of \stepname{Infer DSL}.},
  caption note={\textit{(For parameters involved in the template,
    \PromptParam{RETURN_COUNT} represents the number of runtime return values,
    \PromptParam{BOUND_RULE_VOCABULARY} represents the rule-specific predicates and required Boolean formulas,
    \PromptParam{RULE} represents the original natural-language rule clause,
    \PromptParam{TARGET_DECLARATION} represents the associated entity declaration,
    and \PromptParam{FLATTENED_TARGET_ATTRIBUTES_JSON} represents allowed target-attribute paths and their scalar values.)}},
  label={lst:f2:template},
]
  {Infer DSL}
  {appendices/prompt_artifacts/09_formalize_dsl_translation.template.txt}

\PromptParameter[
  title={\paHeading{Input}{Return count}},
  caption={Example for parameter \PromptParam{RETURN_COUNT} in Listing~\ref{lst:f2:template}.},
  label={lst:f2:param-return-count},
]
  {RETURN_COUNT}
  {appendices/prompt_artifacts/09_formalize_dsl_translation.parameter.return_count.txt}
  
\PromptParameter[
  title={\paHeading{Input}{Bound vocabulary}},
  caption={Example for parameter \PromptParam{BOUND_RULE_VOCABULARY} in Listing~\ref{lst:f2:template}.},
  label={lst:f2:param-vocabulary},
]
  {BOUND_RULE_VOCABULARY}
  {appendices/prompt_artifacts/09_formalize_dsl_translation.parameter.bound_rule_vocabulary.txt}

\PromptParameter[
  title={\paHeading{Input}{Rule}},
  caption={Example for parameter \PromptParam{RULE} in Listing~\ref{lst:f2:template}.},
  label={lst:f2:param-clause},
]
  {RULE}
  {appendices/prompt_artifacts/09_formalize_dsl_translation.parameter.verbatim_clause.txt}

\PromptParameter[
  title={\paHeading{Input}{Target declaration}},
  caption={Example for parameter \PromptParam{TARGET_DECLARATION} in Listing~\ref{lst:f2:template}.},
  label={lst:f2:param-target},
]
  {TARGET_DECLARATION}
  {appendices/prompt_artifacts/09_formalize_dsl_translation.parameter.target_declaration.txt}

\PromptParameter[
  title={\paHeading{Input}{Target attributes}},
  caption={Example for parameter \PromptParam{FLATTENED_TARGET_ATTRIBUTES_JSON} in Listing~\ref{lst:f2:template}.},
  label={lst:f2:param-attributes},
]
  {FLATTENED_TARGET_ATTRIBUTES_JSON}
  {appendices/prompt_artifacts/09_formalize_dsl_translation.parameter.flattened_target_attributes_json.txt}

\LLMResponse[
  title={\paHeading{LLM response}{Infer DSL}},
  caption={Example LLM response for \stepname{Infer DSL} using the prompt template in Listing~\ref{lst:f2:template}.},
  caption note={},
  label={lst:f2:response},
]
  {Infer DSL}
  {appendices/prompt_artifacts/09_formalize_dsl_translation.output.json}

\endgroup

\subsection{EntityLocalizer Supplement: Non-empty Scope Discovery in ERC-721}
\label{sec:app-entity-localizer-supplement}

This subsection supplements EntityLocalizer's \stepname{Infer Scope Regex} example in
Section~\ref{app:entity-localizer}. The ERC-721 document contains named
interfaces that enclose member declarations. The excerpt below shows the
\texttt{ERC721Metadata} interface and one non-empty scope pattern recorded by
the GPT-OSS-20B evaluation run. The named capture group identifies the enclosing
interface, while the opening and closing patterns delimit its body; thus,
\texttt{name}, \texttt{symbol}, and \texttt{tokenURI} can be associated with
\texttt{ERC721Metadata}.

The result is an excerpt of the saved \texttt{hierarchy\_patterns} in
\texttt{target\_description} from evaluation run~1; only the metadata-interface
entry is shown.

\PromptParameter[
  title={\paHeading{Input}{ERC-721 document excerpt}},
  caption={Example for parameter \PromptParam{DOCUMENT_CHUNK} in Listing~\ref{lst:e2:template}: an ERC-721 excerpt containing a named lexical scope and three member declarations.},
  label={lst:e2:erc721-document},
]
  {DOCUMENT_CHUNK (excerpt)}
  {appendices/prompt_artifacts/erc721_scope_document_excerpt.txt}

\LLMResponse[
  title={\paHeading{Recorded result}{ERC-721 scope pattern}},
  caption={One inferred scope pattern retained in the ERC-721 evaluation result.},
  label={lst:e2:erc721-response},
]
  {Infer Scope Regex (supplement)}
  {appendices/prompt_artifacts/erc721_scope.output.json}

\Needspace{16\baselineskip}
\subsection{Baseline Prompts}
\label{sec:app-baseline-prompts}

\begingroup
\PromptSetPlaceholderDelimiters{<<}{>>}

\subsubsection{Spec-in-One (SiO)}

\PromptTemplate[
  caption={SiO: direct specification generation in a single response.},
  caption note={\textit{(\PromptParam{DOCUMENT}: the complete API document, preserving physical source lines.)}},
  label={lst:baseline-sio},
]{Spec-in-One}{appendices/prompt_artifacts/baseline_sio.template.txt}

\Needspace{14\baselineskip}
\subsubsection{Grammar-in-One (GiO)}

\PromptTemplate[
  caption={GiO: joint entity and rule extraction with global grammar generation.},
  caption note={\textit{(\PromptParam{DOCUMENT}: the complete API document;
    \PromptParam{LANGUAGE}: the language hint inferred from the source path.)}},
  label={lst:baseline-gio},
]{Grammar-in-One}{appendices/prompt_artifacts/baseline_gio.template.txt}

\end{document}